\documentclass[a4paper]{jpconf}
\usepackage{graphicx}
\pdfoutput=1
\usepackage{amsmath}
\usepackage{amssymb}
\usepackage{txfonts} %\muup

\begin{document}

\newcommand{\DEG}{\ensuremath{^\circ}}
\newcommand{\DEGC}{\ensuremath{^\circ\mathrm{C}}}
\def\bs{\boldsymbol}

\newlength{\mynull}
\settowidth{\mynull}{0}
\newlength{\mypnull}
\settowidth{\mypnull}{.0}

\newcommand{\w}{\hspace{\mynull}}

\bibliographystyle{iopart-num}

\title{Size-dependent transformation from triangular to rectangular fluxon lattice in Bi-2212 mesa structures}

\author{Holger Motzkau, Sven-Olof Katterwe, Andreas Rydh and Vladimir M Krasnov}

\address{Department of Physics, Stockholm University, AlbaNova University Center, SE-106\,91 Stockholm, Sweden}

\ead{holger.motzkau@fysik.su.se}

\begin{abstract}
We present a systematic study of the field and size dependencies of the static fluxon lattice configuration in Bi-2212 intrinsic Josephson junctions and investigate conditions needed for the formation of a rectangular fluxon lattice required for a high power flux-flow oscillator. 
We fabricate junctions of different sizes from Bi$_{2}$Sr$_2$CaCu$_2$O$_{8+x}$ and Bi$_{1.75}$Pb$_{0.25}$Sr$_2$CaCu$_2$O$_{8+x}$ single crystals using the mesa technique and study the Fraunhofer-like modulation of the critical current with magnetic field. The modulation can be divided into three regions depending on the formed fluxon lattice. At low field, no periodic modulation and no ordered fluxon lattice is found. At intermediate fields, modulation with half-flux quantum periodicity due to a triangular lattice is seen. At high fields, the rectangular lattice gives integer flux quantum periodicity.
We present these fields in dependence on the sample size and conclude that the transitions between the regions depend only on $\lambda_\mathrm{J}(J_\mathrm{c})$ and occur at about $0.4$ and $1.3$ fluxons per $\lambda_\mathrm{J}$, respectively. These numbers are universal for the measured samples and are consistent with performed numerical simulations. 
\end{abstract}

\section{Introduction}

%\cite{Hu10}

Stacked Intrinsic Josephson Junctions (IJJ), naturally formed in $\mathrm{Bi_2Sr_2CaCu_2O}_{8+x}$ (Bi-2212) high-temperature superconductors, are intensely studied candidates for tunable, high-power THz oscillators due to their large energy gap and possibility of power amplification from coupled junctions~\cite{Ozyuzer07,Hu10}.
However, coherent power amplification in flux-flow oscillators requires a rectangular Josephson fluxon lattice. The triangular lattice, which is favored by interplane repulsion of fluxons, results in an out-of-phase oscillation between neighboring junctions which causes destructive interference.
%One of the most important fingerprints of a Josephson junction is the modulation of critical current with magnetic field.
The fluxon lattice has previously been studied from periodic oscillations in the flux-flow resistance~\cite{Ooi02,Kakeya09} and Fraunhofer modulation~\cite{Katterwe09}, and it was shown that the in-phase state can be stabilized by high fields and geometrical confinement in small mesas.
Here we present a systematic study of the field and size dependencies of the static fluxon lattice configuration in Bi-2212 mesa structures.% by means of the appearance of Fraunhofer modulation at low fields, and the disappearance of additional maxima at integer flux values.

A Fraunhofer modulation of the critical current $I_\mathrm{c}$ as a function of magnetic field is one of the fingerprints of the dc-Josephson effect. In the case of IJJ made of Bi-2212, the observed modulation differs from the ordinary Fraunhofer modulation in two ways:
First, the atomic separation between junctions by $s=1.5\,\mathrm{nm}$ results in a short Josephson penetration depth %$\lambda_\mathrm{J}$ which follows from a common normalization~\cite[equation~6-17]{Tin75}, and can be calculated from 
\begin{equation*}
\label{eqn:JosephsonPenetrationDepth}
\lambda_\mathrm{J}=\sqrt{\frac{\Phi_0 s}{4\pi \mu_0 J_\mathrm{c}\lambda_\mathrm{ab}^2}}.
\end{equation*}
Here, $\lambda_\mathrm{ab} \simeq 200\,\mathrm{nm}$ is the London penetration depth.
For typical $c$-axis critical current densities of $J_\mathrm{c}\approx 10^3\,\mathrm{A\,cm^{-2}}$ for Bi-2212 and $10^4\,\mathrm{A\,cm^{-2}}$ for Bi(Pb)-2212~\cite{Katterwe10}, the Josephson penetration depth becomes $0.7\,\mathrm{\muup m}$ and $0.2\,\mathrm{\muup m}$, respectively. Therefore, the junctions are completely shielded at small fields. At slightly higher fields, individual, weakly interacting fluxons enter the junctions. The ordinary Fraunhofer modulation requires a uniform penetration of magnetic field into the junctions. At low fields, therefore, only a decrease in critical current due to motion of disordered fluxons is observed.
%
%At higher fields, vortices start to interact repulsively with each other and with the edge of the mesa (in-plane repulsion) and start to get into an ordered lattice state where modulation is observed. why????
%
 %From the Josephson penetration depth, also an anisotropy parameter~\cite{Koshelev02} 
%???~\cite{Koshelev02}???
%
%
Second, the fluxons distribute into a two-dimensional lattice rather than a one-dimensional chain due to additional interlayer repulsion from fluxons in neighboring junctions. %, and a stable lattice leads to to a higher critical current.
% Depending on the ratio between interlayer and in-plane repulsion, the lattice can obey a triangular or rectangular order. 
At low fields, the interlayer repulsion energetically favors the formation of the triangular lattice, which is stable under high currents for all integer and half-integer flux values.
For higher fields, the rectangular lattice can be formed due to a higher interlayer repulsion and interaction with the boundaries. This lattice withstands a higher current at $\Phi=(k+1/2)\Phi_0$, where $k$ is an integer, similar to the ordinary Fraunhofer modulation.
The rectangular lattice leads to a modulation with an integer flux quantum periodicity as in the single junction case. The triangular fluxon lattice instead has an effectively doubled lattice constant causing a half integer modulation that corresponds to ``adding one flux quantum per two junctions''~\cite{Koshelev02}, and sub-dominant maxima appear.

%why are the fluxons pinned at all, how?

\section{Experimental}

Intrinsic Josephson Junctions of different sizes down to 1.5 times the Josephson penetration depth $\lambda_\mathrm{J}$ were fabricated on top of freshly cleaved single crystals of Bi-2212 and $\mathrm{Bi_{1.75}Pb_{0.25}Sr_2CaCu_2O}_{8+x}$ [Bi(Pb)-2212] using the mesa technique, Ar ion milling and Focused Ion Beam trimming.
% Thereto, a thin protection layer of gold was deposited on the freshly cleaved crystals, and rectangular masks with dimensions down to $500\,\mathrm{nm}$ have been made using optical lithography and/or ion beam induced deposition of platinum ???ref???, followed by an argon ion etching step which determines the number of intrinsic junctions. The resulting mesas still have the gold/platinum with a good electrical contact on top, while the unprotected etched surface quickly passivates. Finally, a planarization layer and gold electrodes are made using photo lithography. %A resulting sample is shown in figure~???, and
Details of the sample fabrication can be found in Ref.~\cite{Krasnov05}.

All measurements were performed in a cryogen-free cryostat at a base temperature of $1.8\,\mathrm{K}$ and magnetic fields up to $17\,\mathrm{T}$. One of the main experimental challenges was the precise alignment of the superconducting copper-oxide planes parallel to the field in order to prevent Abrikosov vortices from entering. This was achieved using a rotational sample holder with a resolution better than $0.02\DEG$ and searching for the maximum of the quasiparticle resistance, indicating the absence of Abrikosov vortices. %After this procedure, the junctions were warmed up above $T_\mathrm{c}$ in to 
Reproducibility of the modulation patterns during field sweeps up and down proves that no Abrikosov vortices are being trapped.
Due to a partial passivation of the top junction, an additional voltage from a non-linear contact resistance arises. This was carefully extracted from the current-voltage characteristics (IVC), fitted, and subtracted.

%To observe this effect, the magnetic field has to be applied strictly parallel to the ab-planes. Otherwise, Abrikosov vortices may be trapped in the superconducting planes and interact with the Josephson vortices and the modulation is disturbed. Therefore, the sample has to be aligned carefully with respect to the magnetic field. It turned out that using the magneto resistance, in particular maximizing the gap voltage, under rotation of the sample at fields between 5 and $10\,\mathrm{T}$ is a reliable alignment method. The stepping motor of the sample holder has a resolution of about, and the achieved alignment was roughly $\pm 0.1\DEG$. After alignment, the sample was heated up above the critical temperature and cooled at zero field to get rid of possible trapped Abrikosov vortices.

Numerical simulations following the coupled sine-Gordon formalism were performed to determine the lattice state at currents just below to the critical current. The average of the normalized displacement of fluxons in neighboring junctions was calculated for different values of $\Phi$ and $L$. Details of the simulation method can be found in the supplementary material of Ref.~\cite{Katterwe09}.

\section{Results and discussion}

\begin{figure}
\centering
\includegraphics[width=5.25cm]{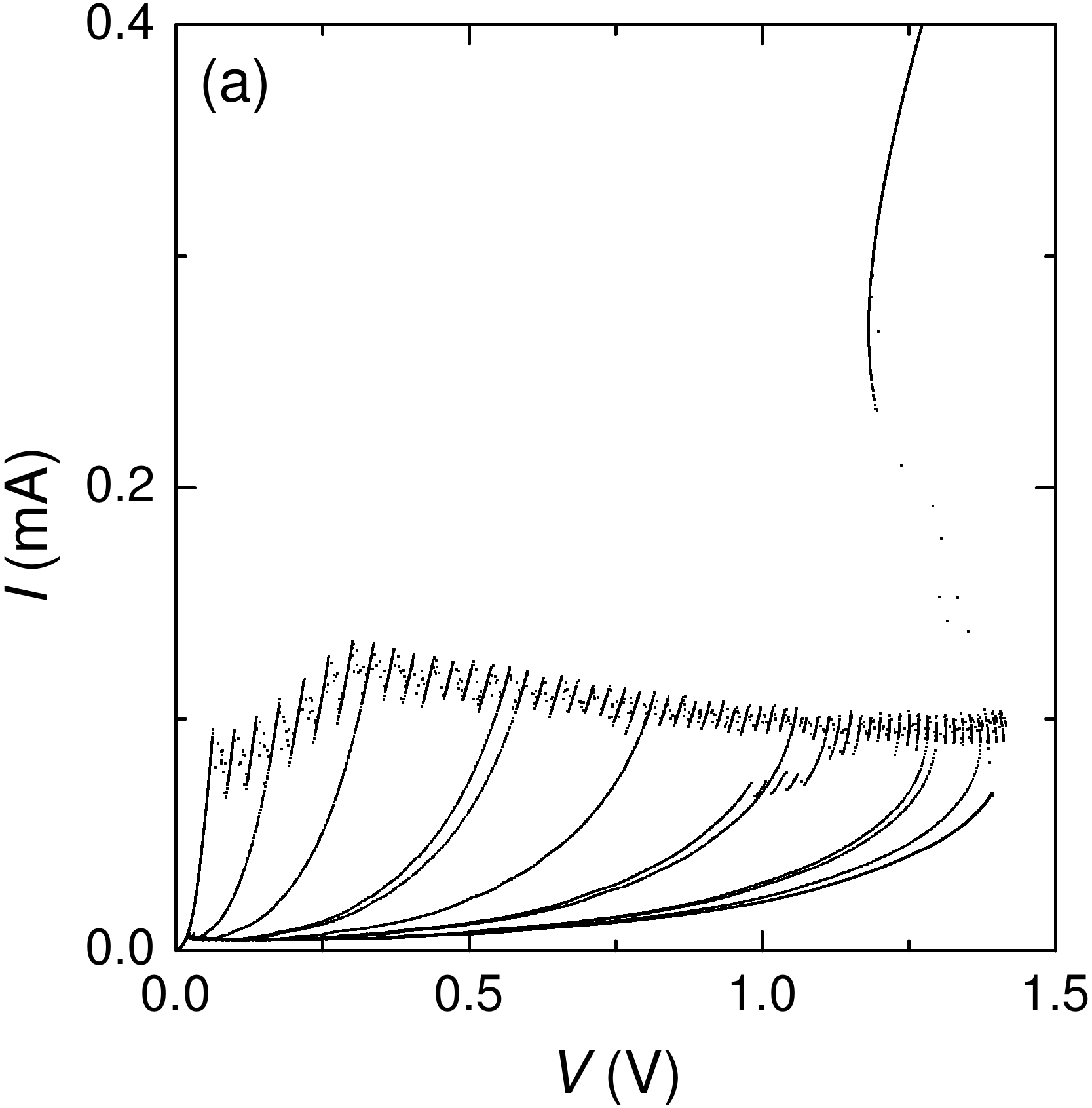}~
\includegraphics[width=4.85cm]{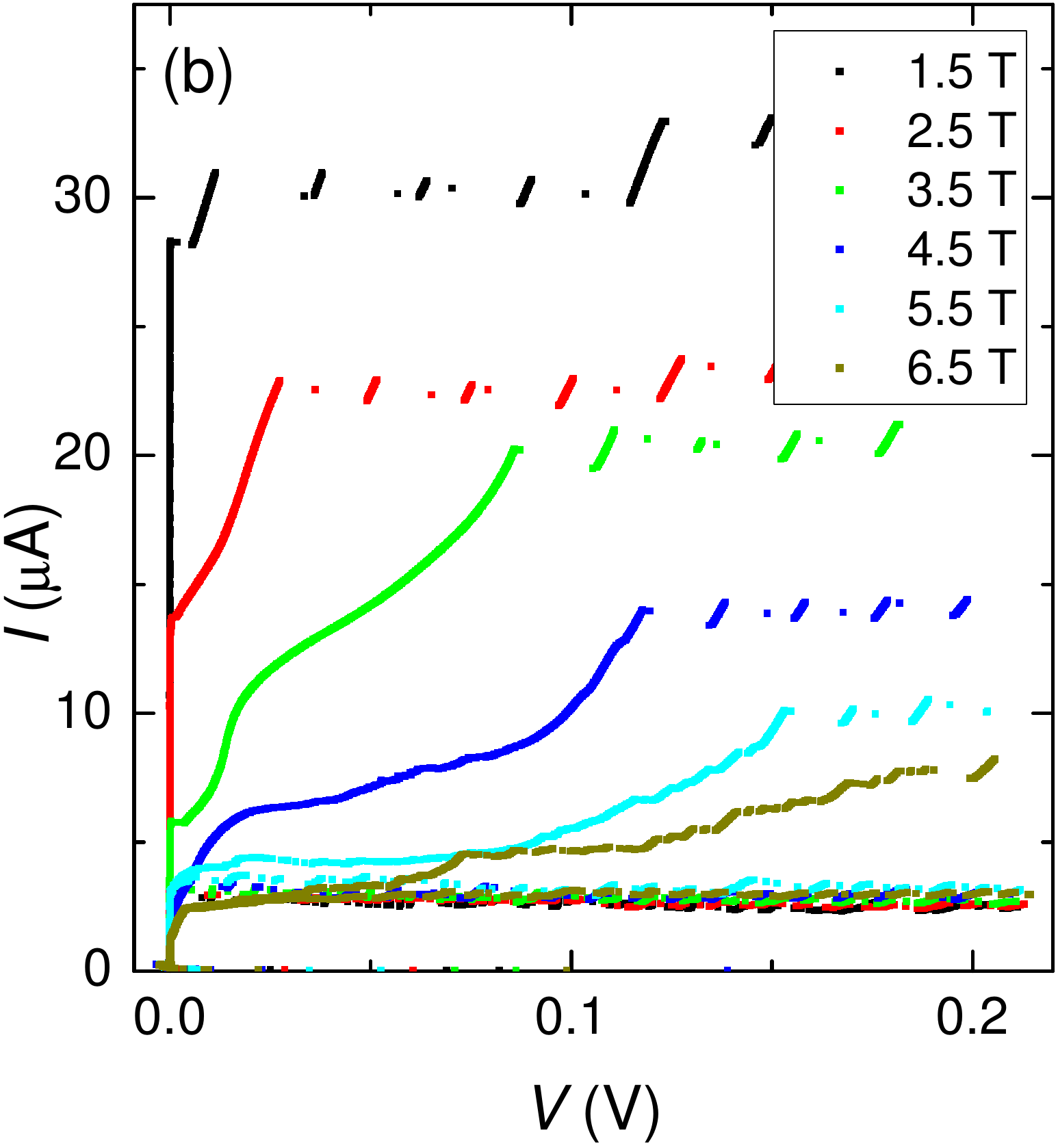}
~
\begin{minipage}[b]{12.6pc}
\caption{\label{fig:ivs} (a) Current-voltage characteristics of a stack with $N=56$ Bi(Pb)-2212 junctions of size $1.2\times2.0\,\mathrm{\muup m^2}$ at $T\approx 1.8\,\mathrm{K}$ and in zero field. (b) The same sample at different magnetic fields with the voltage of the passivated top junction subtracted. A modulating critical current as well as an increasing flux-flow voltage can be observed.
}
\end{minipage}
\end{figure}
\begin{figure}
\centering
\includegraphics[width=7.5cm]{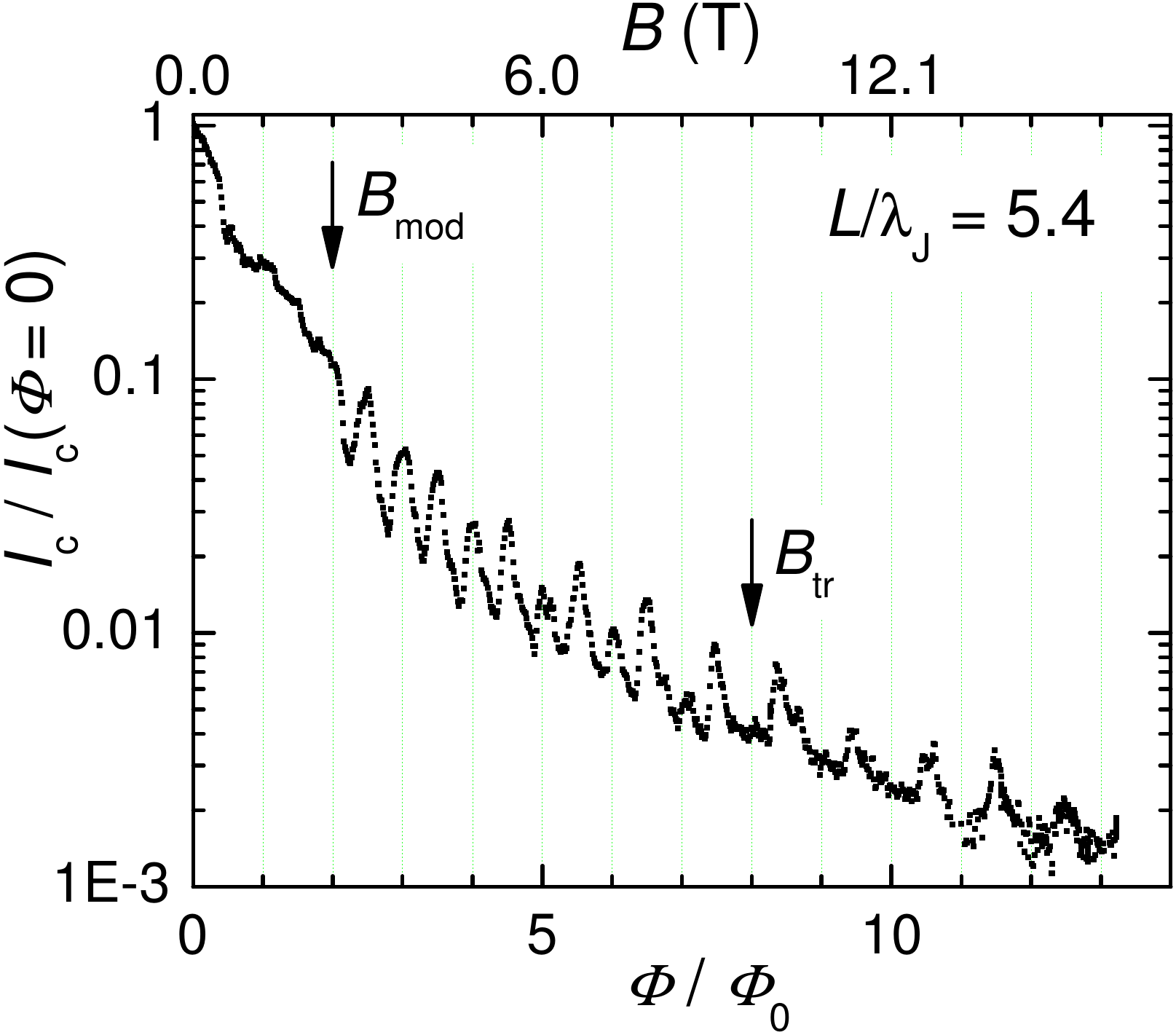}
\hspace{1pc}
\begin{minipage}[b]{18.5pc}
\caption{Fraunhofer-like modulation of the critical current obtained by tracing the maximum current around $V=0$ in figure~\ref{fig:ivs}. The system behavior can be divided into three regions: At low fields, below $B_\mathrm{mod}\approx 2.5\,\mathrm{T}$, no modulation is observed. Above, a modulation with half flux quantum periodicity and dominant as well as sub-dominant maxima arises. At about $B_\mathrm{tr}\approx 9.6\,\mathrm{T}$ this behavior is replaced by a modulation with integer periodicity showing only dominant maxima.}
\label{fig:fraunhofermodulation1}
\end{minipage}
\end{figure}

An IVC of a stack of 56 IJJ at zero field is shown in Fig.~\ref{fig:ivs}(a).
The critical current was extracted automatically from the maximum current within a small window around $V=0$ as shown in figure~\ref{fig:ivs}(b) for different applied fields.
Its dependence on magnetic field is shown in Fig.~\ref{fig:fraunhofermodulation1}.
The field is expressed as flux per junction, $\Phi= B s L$, where $L$ is the mesa size perpendicular to the magnetic field.
%Determining the modulation period the length of the junction can be obtained precisely using the constants $\Phi_0$ and $s$.
%
%The observed modulation is shown in Fig.~\ref{fig:fraunhofermodulation1}. 
The observed behavior differs qualitatively from a simple Fraunhofer pattern in the expected way: At low fields, $B<2.5\,\mathrm{T}$, no modulation is observed, but only a continuous decrease of the critical current. Then, at slightly higher fields, the modulation starts, but obeys extra sub-dominant maxima at fields corresponding to integer numbers of flux quanta in each junction which leads to a half flux quantum periodicity. Finally, at fields higher than of $9.6\,\mathrm{T}$, the sub-dominant maxima disappear and the ordinary $\Phi_0$ periodicity arises.

%Instead of having minimal critical current at fields corresponding to an integer number of flux quanta $\Phi_0$ per junction and dominant maxima in between, the pattern consists of tree different regions.

\begin{figure}
\centering
\includegraphics[width=11cm]{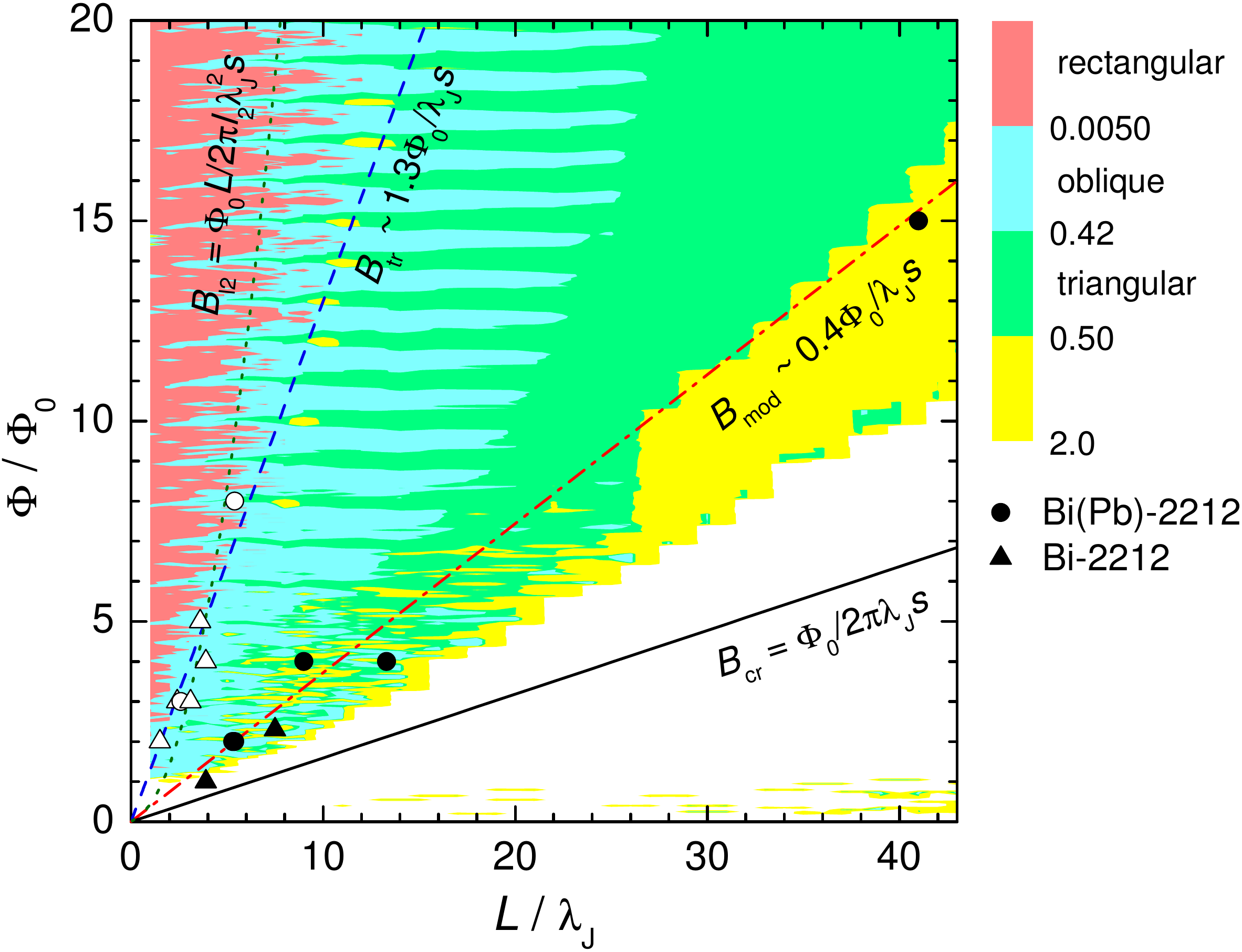}
%\hspace{0.2pc}
%\begin{minipage}[b]{18pc}\hspace{1pc}
%\begin{minipage}[b]{15.8pc}
\caption{Fluxon phase diagram for stacked Josephson junctions:
%triangular symbols are Bi-2212, round symbols Bi(Pb)-2212 junctions.
Starting point of $I_\text{c}(B)$ modulation (filled symbols) and transition from half-integer to integer $\Phi_0$ modulation (open symbols) occur at about 0.4 fluxons per $\lambda_\text{J}$ (red dash-dotted line) and 1.3 fluxons per $\lambda_\text{J}$ (blue dashed line), respectively.
The critical field $B_\text{cr}$ indicating the onset of a homogeneous filling of the layers (solid black line) and the field $B_{l2}$ where a stable rectangular lattice up to the critical current appears (green dotted line) are taken from Ref.~\cite{Koshelev07} and shown for comparison. The background colors illustrate the average of the normalized displacement
of fluxons in neighboring junctions obtained from numerical simulations of the lattice configuration at currents close to the critical current. %: a value close to zero indicates a rectangular (red) and 0.5 a triangular lattice. Intermediate values represent an oblique lattice (blue), while values $>0.5$ indicate a state with irregular ordered and missing vortices or an oblique lattice with a lattice constant of more than two junctions.

%The plot contains the flux of the begin of the modulation $\Phi_\mathrm{mod}$ and of the transition between triangular and rectangular lattice $\Phi_\mathrm{tr}$ against the normalized junction length for the mesas on sample S23 and P22 as well as the data published in~\cite{Katterwe09} where the same crystals as for the S-batch were used. Additionally, data obtained from the simulation program introduced in the same publication is added. Three different regions, namely no regular lattice, a triangular lattice and a rectangular lattice, can be distinguished from the length of the mesa and the flux per junction and collapse for both types of crystal. The border between two regions depends linearly on the size of the mesa, and its slope has the units flux quanta $\Phi_0$ per unit length $\lambda_\mathrm{J}$ corresponding to a constant field which depends on $\lambda_\mathrm{J}$. For illustration, also the line corresponding to one flux quantum per unit length was plotted as a dashed line.
}%(denoted by ``ref'') 
\label{fig:fieldsizeplot}
%\end{minipage}
\end{figure}

The onset of the modulation as well as the transition in periodicity, i.e., the disappearance of additional peaks, was determined for different samples on different crystals, and plotted versus the normalized junction length $L/\lambda_\text{J}$ in Fig.~\ref{fig:fieldsizeplot}. This figure represents the phase diagram of the fluxon lattice in stacked IJJs. As indicated by the red dash-dotted line, the modulation starts at about 0.4 fluxons per $\lambda_\text{J}$, corresponding to 0.8 and $2.5\,\mathrm{T}$ for Bi-2212 and Bi(Pb)-2212, respectively. This is about double the value of the critical field $B_\text{cr}=\Phi_0/2\pi\lambda_\text{J}s$ where fluxons start to fill all layers homogeneously with one fluxon per $2\pi\lambda_\text{J}$~\cite{Koshelev07}. 
The triangular-to-rectangular transition occurs at about $1.3\Phi_0/\lambda_\text{J}$ for the investigated samples, which corresponds to 2.7 and $8.3\,\mathrm{T}$, respectively, for Bi and Bi(Pb).
%the first appearance of the rectangular lattice in the ground state without current occurs if $B > B_\text{cr}/l_1 \cdot L/\lambda_\text{J}$, $l_1\approx 0.675$, and
According to Ref.~\cite{Koshelev07}, the rectangular lattice remains stable up to the critical current for $B > B_\text{cr}/l_2 \cdot L/\lambda_\text{J}$, where $l_2\approx 0.484$. This is plotted as a green dotted curve in Fig.~\ref{fig:fieldsizeplot}. The difference between the linear approximation and the theoretical parabola has little practical significance, since it may be impossible to align large junctions perfectly, while small junctions are in the in-phase state almost as soon as modulations start.

%In experiment, the deviation from the linear approximation plays a minor role since very short junctions obey a stable rectangular lattice already from the beginning of the modulation, while it is practically impossible to measure big junctions up the lattice transition without trapping Abrikosov vortices which disturb the fluxon lattice. 
 %Indeed, the linearity is just an approximation of a quadratic behavior in the range of our investigated samples: .%, and fit well with the experimental values.

The background color of Fig.~\ref{fig:fieldsizeplot} represents the results of our numerical simulations. A number close to zero results from a rectangular lattice (red), while a triangular lattice gives a value of 0.5 (green). Intermediate states with an oblique lattice are blue, while missing fluxons or an oblique lattice with a lattice constant of more than two junctions result in numbers greater than 0.5 and are colored yellow. In the white low-field region, not all junctions are filled with fluxons.
The simulations are consistent with both theory and our experimental data: At low fields, single fluxons enter the junctions and start to form a lattice, which becomes more and more regular until the critical current starts to oscillate with half-integer periodicity. At this size-independent field, the lattice is triangular, but with increasing field the lattice becomes distorted into an oblique lattice near half-integer flux values of $\Phi=(k+1/2)\Phi_0$. At even higher fields the lattice finally becomes rectangular and stabilizes. %while it is still triangular for flux values of $k\Phi_0$. The stabilization of the rectangular lattice occurs slowly around $\Phi=(k+1/2)\Phi_0$,
The triangular lattice at $\Phi=k\Phi_0$, however, never completely disappears~\cite{Koshelev07}.

\section{Summary and conclusion}

We experimentally investigated the fluxon phase diagram of stacked IJJ by studying the Fraunhofer modulation of the critical current in well-aligned small Bi-2212 mesa structures. In particular, we find that the in-phase fluxon state required for the realization of a coherent THz flux-flow oscillator is achieved if $\Phi \gtrsim 1.3 \Phi_0 L/\lambda_\text{J}$ and $\Phi \ne k\Phi_0$. For Bi-2212 and Bi(Pb)-2212, this corresponds to about 3 and 9\,T, respectively.

%in-phase

\ack
Financial support from the Swedish Research Council, the SU-Core Facility in Nanotechnology, and the K.\&A. Wallenberg foundation is gratefully acknowledged.

\section*{References}

\bibliography{../../../MA/literature}        % qhe.bib is the name of our database

%\begin{thebibliography}{9}
%\bibitem{iopartnum} IOP Publishing is to grateful Mark A Caprio, Center for Theoretical Physics, Yale University, for permission to include the {\tt iopart-num} \BibTeX %package (version 2.0, December 21, 2006) with  this documentation. Updates and new releases of {\tt iopart-num} can be found on \verb"www.ctan.org" (CTAN). 
%\end{thebibliography}

\end{document}